\documentclass{elsart}
\usepackage{amsfonts}
\usepackage{amssymb}
\usepackage{amsmath}
\usepackage[dvips]{graphicx,psfrag,overpic,color}

\begin{document}

\begin{frontmatter}

\title{Breaking projective chaos synchronization secure
 communication using filtering and generalized synchronization}
\author[Spain]{G. \'{A}lvarez\corauthref{corr}},
\author[China]{Shujun Li},
\author[Spain]{F. Montoya},
\author[Spain]{G. Pastor} and
\author[Spain]{M. Romera}

\corauth[corr]{Corresponding author: Email: gonzalo@iec.csic.es}

\address[Spain]{Instituto de F\'{\i}sica Aplicada, Consejo Superior de
Investigaciones Cient\'{\i}ficas, Serrano 144---28006 Madrid,
Spain}
\address[China]{Department of Electronic Engineering, City University
of Hong Kong, Kowloon, Hong Kong SAR, China}

\begin{abstract}
This paper describes the security weaknesses of a recently
proposed secure communication method based on chaotic masking
using projective synchronization of two chaotic systems. We show
that the system is insecure and how to break it in two different
ways, by high-pass filtering and by generalized synchronization.

\end{abstract}

\end{frontmatter}
\bibliographystyle{elsart-num}
\sloppy

\section{Introduction}

In recent years, a considerable effort has been devoted to extend
the chaotic communication applications to the field of secure
communications. The possibility of synchronization of two coupled
chaotic systems was first shown by Pecora and Carrol
\cite{pecora90,pecora91,Carroll91} and opened the possibility of
using the signals generated by chaotic systems as carriers for
analog and digital communications. This discovery soon aroused
great interest as a potential means for secure communications
\cite{Yang04}. Accordingly, a great number of cryptosystems based
on chaos have been proposed
\cite{Cuomo93a,Cuomo93b,Wu93,Lozi93,Li01d}, some of them
fundamentally flawed by a lack of robustness and security
\cite{short94,zhou97a,alvarez03,alvarez03b,Li03,Li03a,alvarez04}.

Projective synchronization (PS) is an interesting phenomena
firstly described by Mainieri and Rehacek \cite{Mainieri99}, it
consists in the synchronization of two partially linear coupled
chaotic systems, master and slave, in which the amplitude of the
slave system is a scalar multiple, called scaling factor, of that
of the master system in the phase space. The original study was
restricted to three-dimensional partially linear systems. Later Xu
and Li \cite{Xu02b} showed that PS could be extended to general
classes of chaotic systems without partial linearity, by means of
the feedback control of the slave system; they illustrated the
applicability to Lorenz, Chua and the hyperchaotic R\"{o}ssler
systems.

In a recent paper Li and Xu \cite{Li04} proposed a secure
communication scheme based on PS chaotic masking. The authors
claimed that the unpredictability of the scaling factor of the PS
can additionally enhance the security of communications.
Furthermore, the authors proposed the use of an invertible
function $F$, in such a way that the transmitted ciphertext signal
will be $U(t)=x_1+F[x_1,y_1,z_1,m(t)]$, where $x_1, y_1$ and $z_1$
are the variables of a three-dimensional chaotic system and $m(t)$
is the plaintext (which in their paper is assumed to be a sound
signal). They claimed that the security of the information can
also be guaranteed because the function $F$ could be arbitrarily
chosen. They illustrated the feasibility of the scheme with two
examples, based on the Lorenz and the hyperchaotic R\"{o}ssler
systems, respecitvely.

In this article we show that the proposed cryptosystem is insecure
and how to break it in two different ways, by high-pass filtering
and by generalized synchronization, for both examples based on the
Lorenz and the hyperchaotic R\"{o}ssler systems.

In \cite{Li04}, the first example is based on the following Lorenz
system:
\begin{align}\label{receiver}
  \dot {x_1}  &= \sigma(y_1-x_1),\\
  \dot {y_1}  &=(\mu-z)x_1-y_1,\\
  \dot {z_1}  &=x_1y_1-\rho z_1.
\end{align}
with parameter values $\{\sigma, \mu,\rho\}=\{10, 60, 8/3\}$. The
transmitted signals from the sender to the receiver end are the
shared scalar variable $z_1$ and the ciphertext
$U(t)=x_1+F[x_1,y_1,z_1,m(t)]$, where the function $F$ was
specified as $F[x_1,y_1,z_1,m(t)]=y_1+m(t)$.

The second example is based on the hyperchaotic R\"{o}ssler system
defined by the authors as:
\begin{align}\label{receiver}
    \dot {x_1}  &= -y_1 - z_1,\\
    \dot {y_1}  &=  x_1 + a\,y_1 + w_1,\\
    \dot {z_1}  &= b + x_1z_1,\\
    \dot {w_1}  &= c\,z_1 + d\, w_1.
\end{align}
with parameter values $\{a, b, c, d\}=\{0.25, 3, -0.5, 0.05\}$. In
this example, the transmitted signals from the sender to the
receiver end are the shared scalar variable $w_1$ and the
ciphertext $U(t)=x_1+F[x_1,y_1,z_1,w_1, m(t)]$, where $F$ was
specified as $F[x_1,y_1,z_1,w_1,m(t)]=y_1+m(t)$.

\section{Loose system key specification}

Although the authors seem to base the security of its
communication system on the chaotic behavior of the output of a
chaotic or hyperchaotic non-linear system, no analysis of security
was included. Instead, an unproved assertion saying that ``the
security of information can be guaranteed" was given in the
conclusion.

The first issue to be considered is the key of the system. A
cryptosystem cannot exist without a key, otherwise, it might be
considered as a coding system, but never regarded as a secure
system. In \cite{Li04} it is not considered whether there should
be a key in the proposed system, what it should consist of, what
the available key space would be, and how it would be managed.

When cryptanalyzing a cryptosystem, the general assumption made is
that the cryptanalyst knows exactly the design and working of the
cryptosystem under study, i.e., he knows every detail about the
ciphering algorithm, except the secret key. This is an evident
requirement in today's secure communications systems, usually
referred to as Kerchoff's principle \cite{Stinson95}.

In \cite{Li04} it was stated that the arbitrary selection of the
function $F$ will warrant the information security. But according
to the Kerchoff's principle, the function $F$ must be publicly
known and may not be considered part of the key. At most, its
structure could contain some factors or constants whose values can
play the role of secret key; but, unfortunately, the authors of
\cite{Li04} have not considered such possibility, nor which
conditions the function $F$ might satisfy, nor how many usable
functions there are, nor how much they can contribute to the
system security. Much care must be exercised when selecting the
function $F$. Otherwise, choosing different functions might create
different ciphertexts, that can be decrypted using the same
algorithm though, as shown in Sec.~\ref{sec:recovery}.

\section{Inefficiency as a masking system}
It is supposed that chaotic masking is an adequate means for
secure transmission, because chaotic systems present some
properties as sensitive dependence on parameters and initial
conditions, ergodicity, mixing, and dense periodic points. These
properties make them similar to pseudorandom noise
\cite{Devaney92}, which has been used traditionally as a masking
signal for cryptographic purposes. A fundamental requirement of
the pseudorandom noise used in cryptography is that its spectrum
should be infinitely broad, flat and of much higher power density
than the signal to be concealed. In other words, the plaintext
power spectrum should be effectively buried into the pseudorandom
noise power spectrum. The cryptosystem proposed in \cite{Li04}
does not satisfy this condition. On the contrary, the spectrum of
the signal generated by the Lorenz oscillator is of narrow band,
decaying very fast with increasing frequency, showing a power
density much lower than the plaintext at plaintext frequencies.

\begin{figure}[tb]
\begin{center}
\includegraphics[scale=1]{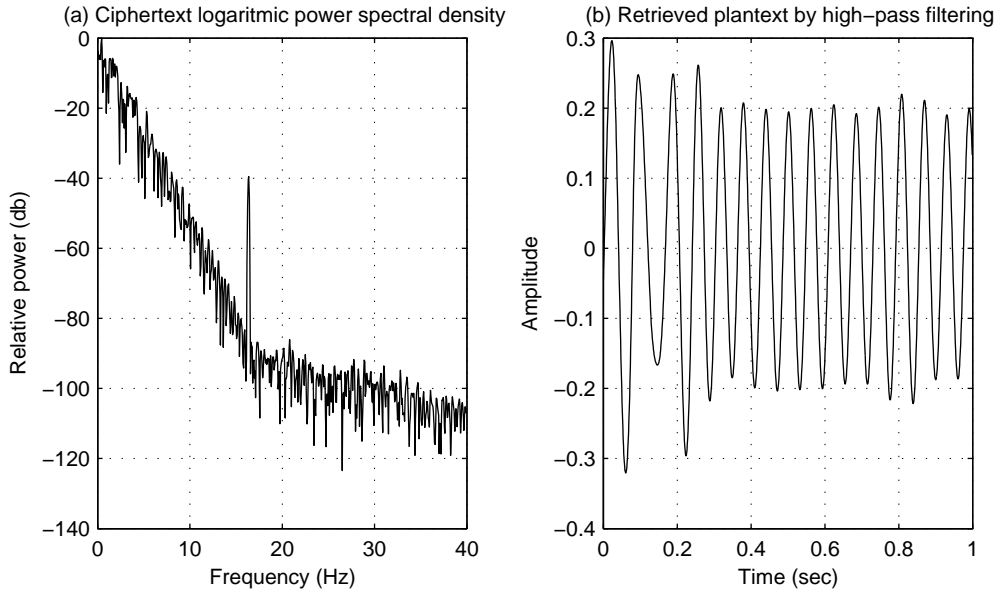}
\caption{Encrypted transmission of a plaintext of amplitude 0.2
and frequency 16.352 Hz, by masking with the Lorenz system
described in \cite{Li04}: (a) logarithmic power spectrum of the
ciphertext; (b) retrieved plaintext by high-pass filtering of the
ciphertext.} \label{fig:lorenz}
\end{center}
\end{figure}

In \cite{Li04} the sound of a water flow was used as the plaintext
message $m(t)$, but no details are given about its waveform or
power spectrum. From \cite[Fig.~2]{Li04} it can be appreciated
that its amplitude is roughly $0.2$. In our simulation we have
used, instead, a well defined plaintext signal  $m(t)= \sin(2 \,
\pi\, 16.352 \,t)$, which corresponds to a pure tone sound of
16.352 Hz, which is the lowest note generated by a musical
instrument, the $\textrm{C}_0$ of a 32 ft pipe of a pipe-organ
\cite{Olson67,Eargle90} and with the same peak amplitude of
\cite[Fig.~2]{Li04}, namely $0.2$.

\begin{figure}[tb]
\begin{center}
\includegraphics[scale=1]{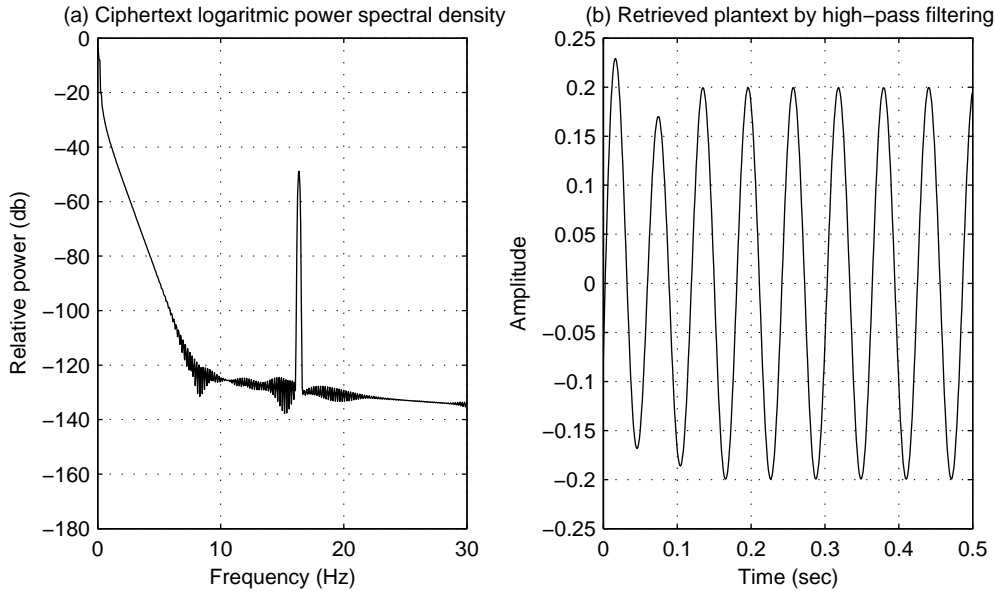}
\caption{Encrypted transmission of a plaintext of amplitude 0.2
and frequency 16.352 Hz, by masking with the hyperchaotic
R\"{o}ssler system described in \cite{Li04}: (a) logarithmic power
spectrum of the ciphertext; (b) retrieved plaintext by high-pass
filtering of the ciphertext.} \label{fig:rosler}
\end{center}
\end{figure}

Figures~\ref{fig:lorenz}(a) and \ref{fig:rosler}(a) illustrate the
logarithmic power spectra of the ciphertext when the Lorenz
attractor and the hyperchaotic R\"{o}ssler attractor are used as
the chaotic system, respectively, with the same parameter values
previously described.

It can be seen that in both examples the plaintext signal clearly
emerges at 16.352 Hz over the background noise created by the
Lorenz and hyperchaotic R\"{o}ssler oscillators, with a power of
$-40~\textrm{db}$ and $-50~\textrm{db}$, respectively, relative to
the maximum power of the ciphertext spectrum, while the power
density of the ciphertext, at neighboring frequencies, falls below
$-80~ \textrm{db}$ and $-125~ \textrm{db}$, respectively.

To recover the plaintext we did not use a chaotic receiver,
instead the ciphertext was high-pass filtered. The procedure is
illustrated in Fig.~\ref{fig:lorenz}(b) and \ref{fig:rosler}(b).
The filter employed was a finite impulse response one. To avoid
phase nonlinearities and distortion, it was constructed with a
512-coefficient Hamming window, with a cutoff frequency of 13 Hz.
The result is a good estimation of the plaintext after a short
initial transient of approximately 0.3 seconds for the Lorenz
system. Note that this is the hardest case an attacker can face
from the point of view of plaintext frequency, because for higher
sound frequencies the spectrum of the background noise created by
the Lorenz oscillator is even lower. While for the hyperchaotic
R\"{o}ssler system the result is a perfect recovery after a short
initial transient of approximately 0.1 seconds.

\section{Generalized synchronization attack}\label{sec:recovery}
The former attack method works only for plaintext frequencies
higher than the 13 Hz cut-off frequency of the high-pass filter
employed. For very low plaintext frequencies the noise created by
the chaotic oscillators effectively masks the plaintext,
preventing its retrieval by direct high-pass filtering. But
plaintext signals of very low frequency may be still retrieved if
we know what kind of non-linear time-varying system was used for
encryption, but without the knowledge of its parameter and initial
condition values. To show such a possibility we have implemented
two cryptanalysis procedures based on generalized synchronization
\cite{Rulkov95,Kocarev96,Yang98j,Alvarez04e}.

\subsection{Breaking the Lorenz system}
To break the PS-based chaotic  masking scheme under study, using
the Lorenz system, we use the following intruder receiver:
\begin{align}\label{receiver}
  \dot {x}_2  &= \sigma^*(y_2-x_2)+ p\, \varepsilon,\\
  \dot {y}_2  &=(\mu^*-z)x_2-y_2+q\,\varepsilon.
\end{align}
where $\{\sigma^*,~\mu^*\}$ are the intruder receiver's Lorenz
system parameters and $\varepsilon$ is the instantaneous value of
the recovery error
$\varepsilon=U(t)-(x_2+y_2)=m(t)+x_1+y_1-x_2-y_2$. The terms $p\,
\varepsilon$ and $q\,\varepsilon$ work as feedback of the recovery
error, in order to achieve generalized synchronization between
sender and receiver; $p$ and $q$ are two scalars that may accept a
wide range of values, from 1 to more than 400, and even one of
them alone may not exist. The synchronism is achieved for any
combination of $p$ and $q$ values, but the amplitudes of $x_2$ and
$y_2$ do not match with those of $x_1$ and $y_1$ while
$\sigma^*\neq\sigma$ or $\mu^*\neq\mu$.

\begin{figure}[tb]
\begin{center}
\psfrag{x1}{$x_1$}\psfrag{x2}{$x_2$}\psfrag{z1}{$z_1$}
\includegraphics[scale=1]{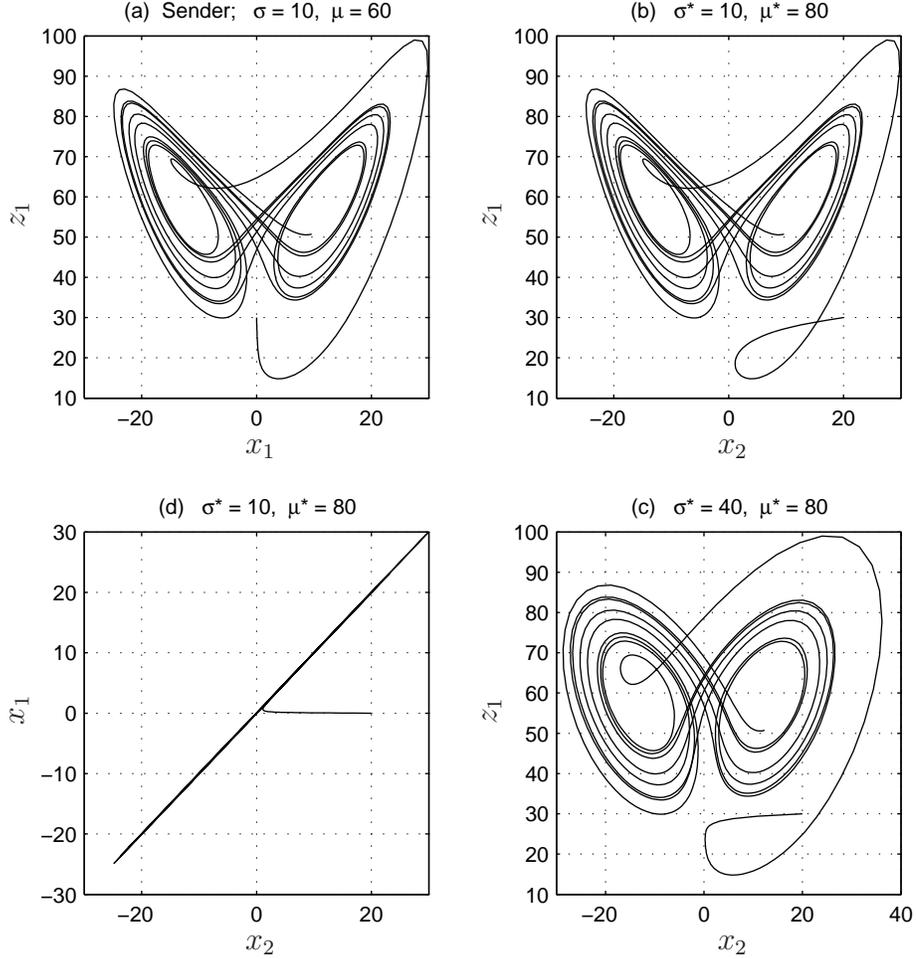}
\caption{Generalized synchronization of the Lorenz attractor: (a)
plot of the sender variables $z_1$ vs. $x_1$, for
$\{\sigma,\mu\}=\{10,60\}$; (b) plot of the  sender variable $z_1$
vs. the intruder receiver variable $x_2$, for
$\{\sigma^*,\mu^*\}=\{10,80\}$; (c) plot of the sender variable
$x_1$ vs. the intruder receiver variable $x_2$, for
$\{\sigma^*,\mu^*\}=\{10,80\}$; (d) plot of the sender variable
$x_1$ vs. the intruder receiver variable $x_2$, when
$\{\sigma^*,\mu^*\}=\{40,80\}$. The initial conditions in all
cases were:
$\{x_1(0),y_1(0),z_1(0),x_2(0),y_2(0)\}=\{0,0.2,30,20,1\}$.}
\label{fig:GSsinchro}
\end{center}
\end{figure}

By making $p=\sigma^*$  Eq. (\ref{receiver}) is simplified,
becoming independent of $y_2$, hence not depending upon the
adjustment of $\mu^*$. In this way $x_1=x_2$ whenever
$\sigma^*=\sigma$, regardless of the value of $\mu^*$. Also we
have found experimentally that the best results for fast
convergence of the synchronism and minimum recovery error are
obtained when $q=\sqrt{\sigma^*}$. With those settings our
intruder receiver is redefined as:
\begin{align}
  \dot {x}_2  &= -2 \sigma^*x_2+ \sigma^*U(t)\label{receiver2a}\\
  \dot {y}_2  &=(\mu^*-\sqrt{\sigma^*}-z)x_2-(1+\sqrt{\sigma^*})
  y_2+\sigma^* U(t).\label{receiver2b}
\end{align}

Figure~\ref{fig:GSsinchro} illustrates the synchronization
mechanism between sender and intruder receiver. The values of the
the sender parameters are $\{\sigma,\mu\}=\{10,60\}$ and the
initial conditions of the sender and receiver are:
$\{x_1(0),y_1(0),z_1(0),x_2(t),y_2(t)\}=\{0,0.2,30,20,1\}$.
Figure~\ref{fig:GSsinchro}(a) shows the first 8 seconds of the
plot of the sender variables $z_1$ vs. $x_1$.
Figure~\ref{fig:GSsinchro}(b) shows the plot of the  sender
variable $z_1$ vs. the intruder receiver variable $x_2$ when
$\{\sigma^*,\mu^*\}=\{10,80\}$; comparing (a) and (b) it can be
seen that both phase portraits are identical, after the short
initial transient originated by the different initial conditions.
Figure~\ref{fig:GSsinchro}(c) shows the plot of the sender
variable $x_1$ vs. the intruder receiver variable $x_2$, when
$\{\sigma^*,\mu^*\}=\{10,80\}$; it can be seen that the phase and
amplitude of $x_2$ match exactly those of $x_1$, after the initial
transient, although $\mu^*\neq\mu$. Finally,
Fig.~\ref{fig:GSsinchro}(d) shows the plot of the sender variable
$x_1$ vs. the intruder receiver variable $x_2$, when $\sigma^*$
and $\mu^*$ completely differ from $\sigma$ and $\mu$, revealing
that both systems are synchronized, although their amplitudes and
phases do not match exactly.

\begin{figure}[tb]
\begin{center}
\begin{overpic}[scale=1]{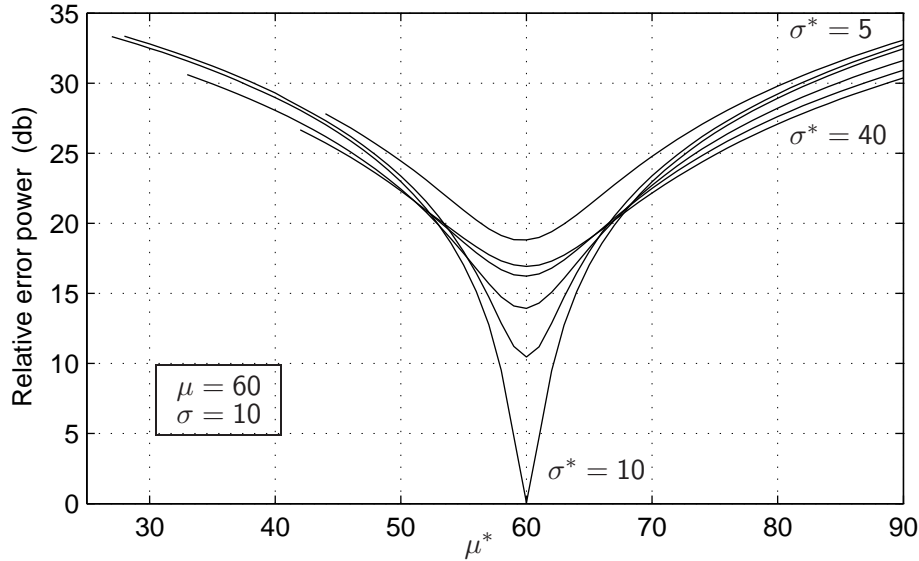}
        \put(85,43){\footnotesize$\sigma^*=\textsf{40}$}
        \put(59,7){\footnotesize$\sigma^*=\textsf{10}$}
        \put(16.82,11.762){\framebox(13.32,7.35){}}
        \put(19,16){\footnotesize$\mu=\textsf{60}$}
        \put(19,13){\footnotesize$\sigma=\textsf{10}$}
        \put(85,54.5){\footnotesize$\sigma^*=\textsf{5}$}
        \put(50,-1){\small$\mu^*$}
        \end{overpic}
\caption{Relative logarithmic representation of the mean of the
error power $\varepsilon^2$, for $\sigma^*=\{5, 7.5, 10, 20, 30,
40\}$ as a function of $\mu^*$. } \label{fig:LogMeanSquaredError}
\end{center}
\end{figure}

We have estimated and recorded the logarithm of the mean value of
the squared error $\varepsilon^2$, i.e. the error power, for the
range of the intruder receiver system parameter values $\sigma^*$
and $r^*$ that give raise to the chaotic behavior of the Lorenz
attractor, with the same transmitter system parameters of the
numerical example presented in \cite[Fig.~2]{Li04} and the
intruder receiver described by Eqs. (\ref{receiver2a}) and
(\ref{receiver2b}). The results are presented in
Fig.~\ref{fig:LogMeanSquaredError}. The mean of $\varepsilon^2$ is
computed along the first 1.5 seconds, after a delay of 0.5 seconds
to let the initial transient finish. It is clearly seen that the
error grows monotonically with the mismatch between the
transmitter and receiver parameters
$\{|\sigma^*-\sigma|,~|\mu^*-\mu|\}$, and that the minimum error
corresponds to the receiver system parameters values $\{\sigma^*,
~\mu^*\}$ exactly matching the transmitter system parameters
values $\{\sigma, ~\mu\}$.

The parameters value recovery procedure consists of the
straightforward search for the minimum recovery error
$\varepsilon$. Once the correct values
$\{\sigma^*,~\mu^*\}=\{\sigma,~\mu\}$ are found, the term
$x_1+y_1-x_2-y_2$ vanishes and the recovery error is just equal to
the plaintext signal $m(t)$.

The search of the correct parameter values $\{\sigma^*,~\mu^*\}$
can be done in the following way: first, select an initial value
for $\sigma^*$ centered in its usable range; second, vary the
value of $\mu^*$ until a minimum error is reached; third, keep
this value and vary the value of $\sigma^*$ until a new minimum
error is reached; four, check if the remaining error $\varepsilon$
is a clean recognizable plaintext, if not repeat the second and
third steps. Note that this method retrieves all at once the
correct values of $\sigma^*$, $\mu^*$, and the plaintext.

\begin{figure}[tb]
\begin{overpic}[scale=1]{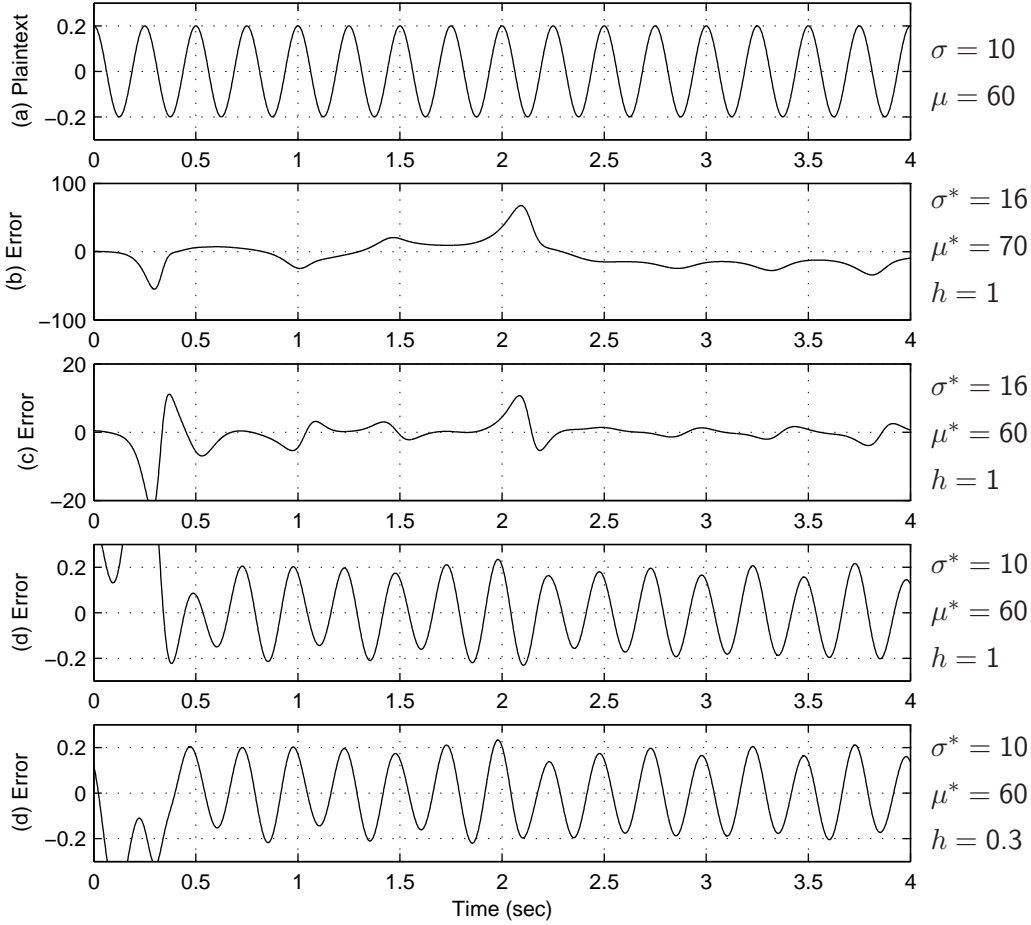}
        \put(100,88){\footnotesize$\mu=\textsf{60}$}
        \put(100,93){\footnotesize$\sigma=\textsf{10}$}
        \put(100,72){\footnotesize$\mu^*=\textsf{70}$}
        \put(100,77){\footnotesize$\sigma^*=\textsf{16}$}
        \put(100,52){\footnotesize$\mu^*=\textsf{60}$}
        \put(100,57){\footnotesize$\sigma^*=\textsf{16}$}
        \put(100,33){\footnotesize$\mu^*=\textsf{60}$}
        \put(100,38){\footnotesize$\sigma^*=\textsf{10}$}
        \put(100,18.5){\footnotesize$\sigma^*=\textsf{10}$}
        \put(100,13.5){\footnotesize$\mu^*=\textsf{60}$}
        \put(100,67){\footnotesize$h=\textsf{1}$}
        \put(100,47){\footnotesize$h=\textsf{1}$}
        \put(100,28){\footnotesize$h=\textsf{1}$}
        \put(100,8.5){\footnotesize$h=\textsf{0.3}$}
        \end{overpic}
\caption{Plaintext and parameter recovery of the Lorenz system by
generalized synchronization analysis; the transmitter parameters
are $\{\sigma,\,\mu,\,\rho\}=\{10, 60, 8/3\}$: (a) plaintext
$m(t)=\cos(2\pi 4 t)$; (b) receiver error for unadjusted intruder
receiver parameters $\{\sigma^*,\,\mu^*\}=\{16, 70\}$; (c)
receiver error for partially adjusted intruder receiver parameters
$\{\sigma^*,\,\mu^*\}=\{16, 60\}$; (d) receiver error for correct
intruder receiver parameters $\{\sigma^*,\,\mu^*\}=\{10,60\}$; (e)
receiver error for correct intruder receiver parameters
$\{\sigma^*,\,\mu^*\}=\{10,60\}$, and unadjusted function
$F=-0.7\,x_1+0.3,y_1+m(t)$.} \label{fig:GSretrieved}
\end{figure}

The procedure is illustrated in Fig.~\ref{fig:GSretrieved}, for a
plaintext $m(t)=\cos(2\pi 4 t)$, whose frequency is so low that it
cannot be retrieved by the previously described direct high-pass
filtering method. When the initial parameter values are chosen at
random as $\{\sigma^*,\,\mu^*\}=\{16, 70\}$, the corresponding
error reaches a peak value near 70. Then the $\mu^*$ value is
varied until a minimum of the error is found for $\mu^*=60$, it
can be seen that now the peak error value after the initial
transient is reduced to about 15. Next, the $\sigma^*$ value is
varied until a new error minimum is reached for $\sigma^*=10$, now
the error is reduced to the plaintext itself, plus some noise of
reduced amplitude that may be easily removed by a high-pass
filter, if necessary.

We have found that our method woks as well for the whole family of
functions of the form $F=(h-1)\,x_1+h\,y_1+m(t)$, were $h$ is a
scalar of any value. The plaintext $m(t)$ is correctly recovered
for any ciphertext of the form $U(t)=x_1+F[x_1,y_1,m(t)]$. This
fact demonstrates that a great care must be exercised when
selecting the function $F$, because some changes in its factors
and constants values may be useless to enlarge the key space,
hence not improving the system security at all. In
Fig.~\ref{fig:GSretrieved}(e) a time story of the retrieved
message for a function of the form $F=-0.7\,x_1+0.3,y_1+m(t)$ is
presented, when decoded by an intruder receiver adjusted to
recover a function of the type $F=y_1+m(t)$. It can be observed
that the only difference with Fig.~\ref{fig:GSretrieved}(d) is the
magnitude and duration of the initial transient.

\subsection{Breaking the hyperchaotic R\"{o}ssler system}
To break the PS-based chaotic masking scheme under study, when the
hyperchaotic R\"{o}ssler system is used to generate the masking
signal, we follow a similar procedure to the one used in the
preceding section. Now the intruder receiver is:
\begin{align}\label{receiver2}
    \dot {x_2}  &= -y_2 - z_2 +  p\,\, \varepsilon,\\
    \dot {y_2}  &= x_2 + ay_2 + w_1,\\
    \dot {z_2}  &= b + x_2z_2.
\end{align}
were $p$ is a scalar. We have found the best results with $p=10$.

Here the instantaneous value of the recovery error is defined as
$\varepsilon=U(t)-(x_2+y_2)=m(t)+x_1+y_1-x_2-y_2$. When the
synchronism is reached it happens that $x_1+y_1=x_2+y_2$, hence
the error is $\varepsilon=m(t)$, thus allowing the exact recovery
of the plaintext.

\begin{figure}[tb]
\begin{overpic}[scale=1]{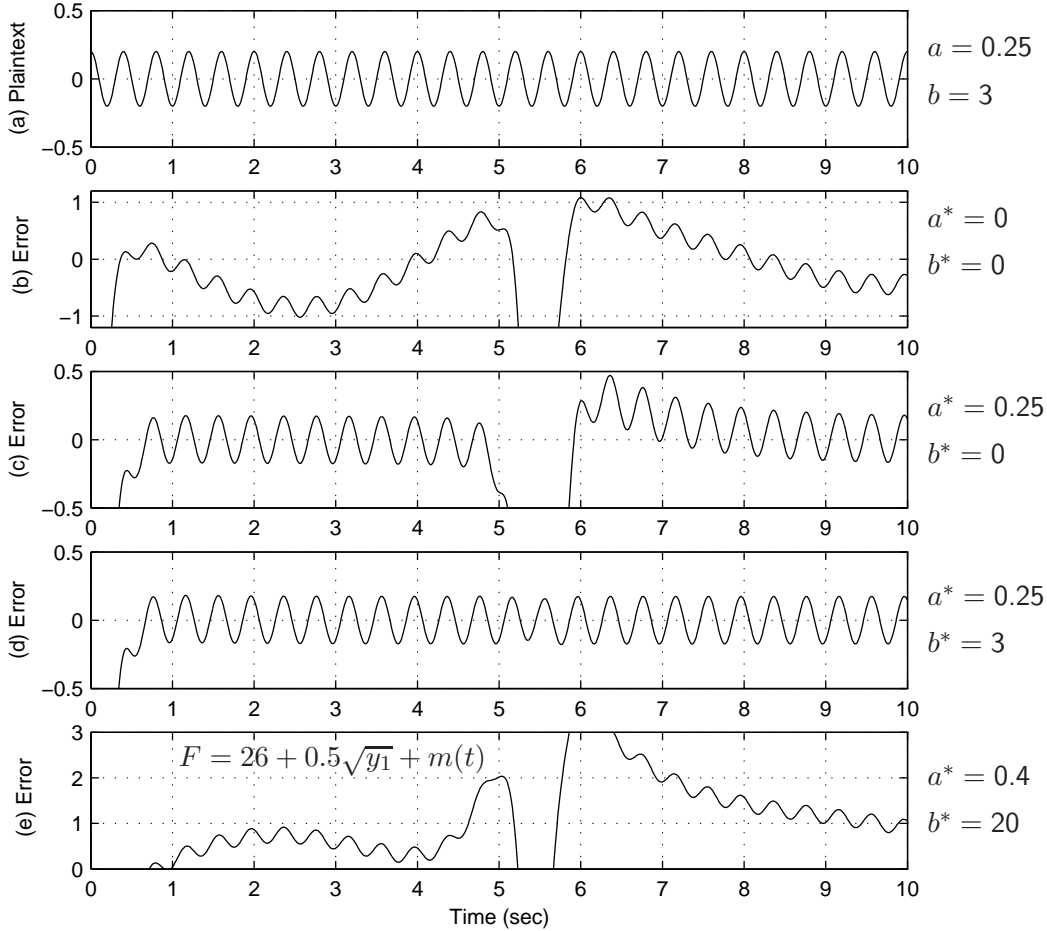}
        \put(98,88){\footnotesize$b=\textsf{3}$}
        \put(98,93){\footnotesize$a=\textsf{0.25}$}
        \put(98,70){\footnotesize$b^*=\textsf{0}$}
        \put(98,75){\footnotesize$a^*=\textsf{0}$}
        \put(98,50){\footnotesize$b^*=\textsf{0}$}
        \put(98,55){\footnotesize$a^*=\textsf{0.25}$}
        \put(98,30){\footnotesize$b^*=\textsf{3}$}
        \put(98,35){\footnotesize$a^*=\textsf{0.25}$}
        \put(98,16){\footnotesize$a^*=\textsf{0.4}$}
        \put(98,11){\footnotesize$b^*=\textsf{20}$}
        \put(19,18){\footnotesize$F=26+0.5 \sqrt{y_1}+ m(t)$}
        \end{overpic}
\caption{Plaintext and parameter recovery by generalized
synchronization analysis; the sender parameters are $\{a, b, c,
d\}=\{0.25, 3, -0.5, 0.05\}$: (a) plaintext $m(t)= \cos(2 \pi 2.5
t)$; (b) receiver error for unadjusted intruder receiver
parameters $\{a^*,\,b^*\}=\{0,0\}$; (c) receiver error for
partially adjusted intruder receiver parameters
$\{a^*,\,b^*\}=\{0.25,0\}$; (d) receiver error for the correct
intruder parameter values $\{a^*,\,b^*\}=\{0.25,3\}$. (e) receiver
error for wrong intruder receiver parameter values
$\{a^*,\,b^*\}=\{0.25,3\}$, and unadjusted function $F=26+0.5
\sqrt{y_1}+ \cos(2\pi 2.5 t)$.} \label{fig:RoslerGS}
\end{figure}

Figure~\ref{fig:RoslerGS} illustrates the plaintext recovered for
various intruder receiver parameter values sets, being the
plaintext $m(t)= \cos(2 \pi 2.5 t)$, whose frequency is so low
that it cannot be retrieved by the previously described direct
high-pass filtering method. It can be observed that the intruder
receiver is quite insensitive to the values of the parameters
$\{a^*,b^*\}$ and the structure of the function $F$. In
Fig.~\ref{fig:RoslerGS}(b) it is shown that when the parameter
values of sender and intruder receiver do not match at all, the
plaintext is still visible, although a residual interference is
present, but its intensity is not big enough to preserve the
confidentiality of the communications. This interference can be
easily removed, if desired, by trial and error in few steps. The
first parameter to be adjusted is $a^*$, because it is the most
influent on the shape of the retrieved waveform. In
Fig.~\ref{fig:RoslerGS}(c) it is shown the waveform for $a^*$
correctly adjusted while $b^*$ is kept unadjusted. Then the
parameter $b^*$ must be adjusted until a clean recovered plaintext
waveform is reached, as illustrated in Fig.~\ref{fig:RoslerGS}(d).

We have tested several different invertible functions $F$ as
building blocks of the sender ciphertext, and it has been observed
that quite different functions allow for the almost correct
retrieving of the plaintext. Figure~\ref{fig:RoslerGS}(e)
dramatically illustrates this fact: when a function as complicated
as $F=26+0.5 \sqrt{y_1}+ m(t)$ is used for transmission, while
maintaining an intruder receiver designed for decoding a function
of the type $F=y_1+m(t)$, together with a total parameter mismatch
between sender and receiver as
$\{a,b,a^*,b^*\}=\{0.25,3,0.4,20\}$, we can see that the retrieved
waveform still retain enough plaintext information to completely
compromise the security of the communication.

\section{Other weaknesses of the proposed system}

The knowledge of the scaling factor $\alpha$ is not required to
retrieve the plaintext, if a high-pass filter attack or a
generalized synchronization recovering procedure are implemented,
as we did.  Hence the scaling factor does not add any strength to
the system security, as opposed to the claims by the authors of
\cite{Li04}.

Thanks to the plain transmission of the shared scalar variables
$z_1$, or $w_1$, inherent to the PS scheme, the third Lorenz
system parameter $\rho$ and the hyperchaotic R\"{o}ssler
parameters $c$ and $d$ need not be determined to recover the
plaintext when using a generalized synchronization receiver, in
opposition to other cryptosystems that make use the Lorenz or
hyperchaotic R\"{o}ssler systems as a masking signal. Therefore an
additional advantage is offered by the authors of \cite{Li04} to
the opponent eavesdropper.

\section{Conclusion}
In summary, the proposed PS-based chaotic masking cryptosystem is
rather weak, since it can be broken in two easy ways, the first
one being ignorant of the transmitter precise structure and the
second one knowing the transmitter structure but ignoring its
parameter values. The alleged security advantage of the
cryptosystem based on the eavesdropper lack of knowledge of the
scaling factor $\alpha$ is incorrect, its knowledge is completely
irrelevant to retrieve the plaintext. The function $F$ does not
clearly enhance the security, owing to the fact that a wide range
of different functions generate ciphertexts equally breakable with
the same intruder receiver with a unique parameter adjustment.
There is no mention about what the key is, nor which is the key
space, a fundamental aspect in every secure communication system.
The total lack of security discourages the use of this
communication scheme for secure applications, unless some
modifications are made to essentially enhance its security.

\ack{This work was supported by Ministerio de Ciencia y
Tecnolog\'{\i}a of Spain, research grants TIC2001-0586 and
SEG2004-02418, and by the Applied R\&D Center, City University of
Hong Kong, Hong Kong SAR, China, under Grants no. 9410011 and no.
9620004.}


\end{document}